\documentstyle[preprint,pra,eqsecnum,aps]{revtex}
\tightenlines

\begin{document} \draft

\title{\Large \bf WIGNER'S PHOTONS}

\author{Y. S. Kim\thanks{electronic mail: yskim@physics.umd.edu}}
\address{Department of Physics, University of Maryland, College Park,
Maryland 20742, U.S.A.}

\maketitle

\begin{abstract}

If Einstein's photon is $E = cp = \hbar\omega$, Wigner's photon is
its helicity which is a Lorentz-invariant concept coming from the
$E(2)$-like little group for massless particles.  In addition, the
$E(2)$-like little group has two translation-like degrees of freedom.
What happens to them?  They are associated with the gauge degree of
freedom.  Since the physics of polarized light waves can be formulated
within the framework of the Lorentz group, it is now possible to use
polarization experiments to study the $E(2)$-like little group in
terms of quantities that can be measured in laboratories.

\end{abstract}

\section{Introduction}
The purpose of this meeting is to discuss photons, and there are many
papers based on Wigner functions.  The oscillator-based Wigner function
is the natural language for the $Sp(2)$ and $Sp(4)$ groups~\cite{knp91}.
These two groups also happened to be the natural languages for one- and
two-mode squeezed states respectively.  Since they are are isomorphic
to $O(2,1)$ and $O(3,2)$ Lorentz groups respectively, the squeezed state
is not only the physics of coherent photon states, but also the physics
of Lorentz transformations.  Let us note also that Lorentz boosts are
squeeze transformations~\cite{dir49}.

Wigner's influence is not limited to Wigner functions and squeezed
states.  In his 1931 book entitled {\it Gruppentheorie und ihre
Anwendung auf Quantenmechanik der Atomspektren}~\cite{wig31}, Wigner
gives a complete group theoretical description of angular-momentum
states.  Of course, transitions between quantized atomic states are
mediated by photons carrying their angular momenta.  It is therefore
safe to say that this book was the first book on the quantum theory of
photons, as well as on applications of group theory in physics.  This
book was translated from German into English and from the left-handed
coordinate system to the right-handed coordinate system by Griffin in
1959~\cite{wig59}.

The main purpose of the present report is to discuss the plane-wave
solution of Maxwell's equations within the framework of Wigner's
little groups.  In his 1939 paper on representations of the
Poincar\'e group~\cite{wig39}, Wigner formulated the internal
space-time symmetries of relativistic particles in terms of the little
groups.  There he observed that the internal space-time symmetry of
massless particles is dictated by a subgroup of the Lorentz group
isomorphic to $E(2)$ or the two-dimensional Euclidean group.  But
Wigner did not address the question of how his formalism can accommodate
Maxwell's equations, and left this as a home-work problem for younger
generations.  This homework problem has now been solved by various
authors, and Wigner's 1939 paper provides a place for photons and
Maxwell's equations.

In 1949, Wigner published a paper on localization of relativistic
systems~\cite{newig49}.  This paper led to the problem known today as
the photon localization~\cite{hkn95}.  As in the case of the little
group, this localization problem with a single photon.  However,
the Lorentz group continues to play
important roles in many-photon systems.  As is well demonstrated at
this conference, the squeezed state of light is the physics of the
$O(2,1)$ and $O(3,2)$ Lorentz groups.  In addition, it has been
found recently that the Lorentz group is the natural language for
the theory of polarized lights which is almost 150 years
old~\cite{stokes52,swind75}.

Indeed, Wigner cannot be separated from classical and modern optics.
In Sec.~\ref{max}, we establish a connection between Wigner's
representation theory and Maxwell's equations.  In Sec.~\ref{others}
we discuss other areas of the science of photons where Wigner made
fundamental contributions.

\section{Maxwell and Wigner}\label{max}

In 1939, Wigner observed that internal space-time symmetries of
relativistic particles are dictated by their respective little
groups~\cite{wig39}.  The little group is the maximal subgroup of the
Lorentz group which leaves the four-momentum of the particle invariant.
The Lorentz group is generated by three rotation generators $J_{i}$
and three boost generators $K_{i}$, which satisfy the commutation
relations:
\begin{equation}\label{comm1}
\left[J_{i}, J_{j}\right] = i \epsilon_{ijk} J_{k} , \quad
\left[J_{i}, K_{j}\right] = i \epsilon_{ijk} K_{k} , \quad
\left[K_{i}, K_{j}\right] = -i \epsilon_{ijk} J_{k} .
\end{equation}
If a massive particle is at rest, it momentum is invariant under
three-dimensional rotations.
Thus, its little group is generated by $J_{1}, J_{2}$, and $J_{3}$,
and its spin orientation is changed under the little group
transformation.  If we use the metric convention $(x, y, z, t)$,
the transformation matrix applicable to spin-1 particles is
\begin{equation}
\pmatrix{r_{11} & r_{12} & r_{13} & 0 \cr
r_{21} & r_{22} & r_{23} & 0 \cr r_{31} & r_{32} & r_{33} & 0 \cr
0  & 0 & 0 & 1} ,
\end{equation}
where the three-by-three matrix consisting of the first three rows
and columns constitute a rotation matrix in the three-dimensional
space.  This matrix leaves the four-momentum $(0, 0, 0, m)$ invariant,
but can change the direction of the angular momentum.

For a massless particle, it is not possible to find a Lorentz frame
in which the particle is at rest.  We can however assume that its
momentum is in the $z$ direction.  Then the momentum is invariant
under the subgroup of the Lorentz group generated by
\begin{equation}
J_{3} , \qquad N_{1} = K_{1} - J_{2} , \qquad N_{2} = K_{2} + J_{1} .
\end{equation}
Wigner noted in his 1939 paper that these generators satisfy the
same set of commutation relations as those for the two-dimensional
Euclidean group consisting of one rotation and translations in two
different directions.  With these generators, we can construct the
following transformation matrices generated by $J_{3}$.
\begin{equation}\label{rotz}
\pmatrix{\cos\phi & 0 & -\sin\phi & 0 \cr 0 & 1 & 0 & 0 \cr
\sin\phi & 0 & \cos\phi & 0 \cr 0 & 0 & 0 & 1 } ,
\end{equation}
which performs rotations around the momentum.  It is not difficult
to associate this matrix with the helicity of the photon, and it is
an invariant concept~\cite{chou58,jacob59,hks85}.
In addition, they generate transformation matrices of the form
\begin{equation}\label{ugly}
\pmatrix{1 & 0 & -u & u \cr 0 & 1 & -v & v \cr
u & v & 1 - (u^{2} + v^{2})/2 & (u^{2} + v^{2})/2 \cr
u & v & -(u^{2} + v^{2})/2 & 1 + (u^{2} + v^{2})/2 } .
\end{equation}
This expression is given in Wigner's original paper, and this matrix
leaves the four-momentum
\begin{equation}\label{4mom}
(0, 0, \omega, \omega)
\end{equation}
invariant.  But its strange appearance kept physicists away from the
matrix for many years.

One quick solution to this problem was to avoid this expression.
For this reason, for many years, there was a tendency to restrict
representations which are invariant under this transformation.
In 1964~\cite{wein64}, Weinberg systematically constructed
representations which are invariant under this transformation, and
ended up with the gauge-invariant electromagnetic tensor.  This
leads to a suspicion that the not-so-nice-looking matrix of
Eq.(\ref{ugly}) could be a gauge transformation.  What else could it
be?

However, our imagination did not reach that quickly.  This observation
was made by several independent research groups after
1970~\cite{janner71,kuper76}.   These days, we have an easy way to
see this~\cite{hks82}.  A plane wave propagating along the $z$ direction
can have a four-potential
$(A_{1}, A_{2}, A_{3}, A_{0})$, with the Lorentz condition
$A_{3} = A_{0}$.  If the third and the fourth components are equal,
the four-by-four matrix of Eq.(\ref{ugly}) becomes
\begin{equation}\label{ugly2}
\pmatrix{1 & 0 & 0 & 0 \cr 0 & 1 & 0 & 0 \cr
u & v & 1  & 0 \cr u & v & 0 & 1 } .
\end{equation}
Furthermore, its application to the four-potential leads to
\begin{equation}\label{ugly3}
\pmatrix{1 & 0 & 0 & 0 \cr 0 & 1 & 0 & 0 \cr
u & v & 1  & 0 \cr u & v & 0 & 1 }
\pmatrix{A_{1} \cr A_{2} \cr A_{3} \cr A_{0}} =
\pmatrix{A_{1} \cr A_{2} \cr A_{3} \cr A_{0}} +
\pmatrix{0 \cr 0 \cr uA_{1} + v A{2} \cr uA_{1} + vA_{2}} .
\end{equation}
This means the application of the matrix of Eq.(\ref{ugly}) to
the four-potential leads to the addition of a four-vector
proportional to the four-momentum of Eq.(\ref{4mom}).  This is
a gauge transformation!

We thus can give a complete interpretation of Wigner's $E(2)$-like
little group for photons in terms of Maxwell's equations.  In 1964,
Weinberg started his analysis starting from the two-by-two
representations of the $SL(2,C)$ group generated by
\begin{equation}\label{sl2gen}
J_{i} = {1 \over 2} \sigma_{i} , \qquad
K_{i} = {i \over 2} \sigma_{i} ,
\end{equation}
which satisfy the closed set of commutation relations given in
Eq.(\ref{comm1}).  The commutation relations remain invariant under
the sign change of the boost generators.  Thus the representations
of the $SL(2,C)$ group are twin representations, and there are
four components in the spinor as in the case of the Dirac spinor.
This leads to sixteen independent components of the tensor, and the
systematic reduction leads to the four-potential and the Maxwell
tensor~\cite{hks86,knp86,baskal97}.

\section{Other Aspects of Wigner's Photons}\label{others}
In order to complete the particle picture of photons, we have to
second-quantize the Maxwell fields.  This eventually leads to
quantum electrodynamics.  However, unlike Schr\"odinger wave functions,
quantum fields do not carry probability interpretation.  This point
was first observed in Wigner's 1949 paper on localization of
relativistic systems~\cite{newig49}.  There has been some progress
in recent years on this question in terms of wavelets, but this
fundamental problem still remains unsolved~\cite{hkn95}.

On the other hand, there has been a concrete progress in recent years
on applications of the Lorentz group in polarization optics.  This
subject deals with the Jones matrix and the Stokes parameters which
are constantly used in laboratories where polarized photons and light
waves are observed.  If a light wave propagates along the $z$
direction, the electric field vector can be written as
\begin{equation}\label{expo1}
\pmatrix{E_{x} \cr E_{y}} =
\pmatrix{A \exp{\left\{i(kz - \omega t + \phi_{1})\right\}}  \cr
B \exp{\left\{i(kz - \omega t + \phi_{2})\right\}}} .
\end{equation}
where $A$ and $B$ are real and positive numbers,
and $\phi_{1}$ and $\phi_{2}$ are the phases of the $x$ and $y$
components respectively. This column matrix is called the Jones vector.

The transformation takes place when the light beam goes through an
optical filter whose transmission properties are not isotropic.
The absorption coefficient in one transverse direction could be
different from the coefficient along the other direction.  Thus, there
is the ``polarization'' coordinate in which the absorption can be
described by
\begin{equation}\label{atten}
\pmatrix{e^{-\eta_{1}} & 0 \cr 0 & e^{-\eta_{2}}} =
e^{-(\eta_{1} + \eta_{2})/2} \pmatrix{e^{\eta/2} & 0 \cr 0 & e^{-\eta/2}}
\end{equation}
with $\eta = \eta_{2} - \eta_{1}$ .
This attenuation matrix tells us that the electric fields are attenuated at
two different rates.  The exponential factor $e^{-(\eta_{1} + \eta_{2})/2}$
reduces both components at the same rate and does not affect the degree of
polarization.  The effect of polarization is solely determined by the
squeeze matrix
\begin{equation}\label{sq1}
S(0, \eta) = \pmatrix{e^{\eta/2} & 0 \cr 0 & e^{-\eta/2}} .
\end{equation}
This type of mathematical operation is quite familiar to us from
squeezed states of light~\cite{knp91}.
Another basic element is the optical filter with two different values
of the index of refraction along the two orthogonal directions.  The
effect of this filter can be written as
\begin{equation}\label{phase}
\pmatrix{e^{i\lambda_{1}} & 0 \cr 0 & e^{i\lambda_{2}}}
= e^{-i(\lambda_{1} + \lambda_{2})/2}
\pmatrix{e^{-i\lambda/2} & 0 \cr 0 & e^{i\lambda/2}} ,
\end{equation}
with $\lambda = \lambda_{2} - \lambda_{1}$ .
In measurement processes, the overall phase factor
$e^{-i(\lambda_{1} + \lambda_{2})/2}$
cannot be detected, and can therefore be deleted.  The polarization
effect of the filter is solely determined by the matrix
\begin{equation}\label{shif1}
P(0, \lambda) = \pmatrix{e^{-i\lambda/2} & 0 \cr 0 & e^{i\lambda/2}} .
\end{equation}
If the polarization coordinate is the same as the $xy$ coordinate where
the electric field components take the form of Eq.(\ref{expo1}), the
above attenuator is directly applicable to the column matrix of
Eq.(\ref{expo1}).  If the polarization coordinate is rotated by an angle
$\theta/2$, or by the matrix
\begin{equation}\label{rot}
R(\theta) = \pmatrix{\cos(\theta/2) & -\sin(\theta/2)
\cr \sin(\theta/2) & \cos(\theta/2)} .
\end{equation}

Repeated applications of squeeze matrices of Eq.(\ref{sq1}) and phase
shifters of the form given in Eq.(\ref{rot}) together with rotation
matrices of the above form will give a six-parameter transformation
matrix of the form~\cite{hkn96}
\begin{equation}
\pmatrix{\alpha & \beta \cr \gamma & \delta} ,
\end{equation}
generated by the two-by-two matrices given in Eq.(\ref{sl2gen}).  This
is how the Jones-matrix formalism is framed into the Lorentz group.
As we mentioned at the end of Sec.~\ref{max}, it is possible to
construct a four-vector from the four $SL(2,C)$ spinors.  The element
of this four-vector are the Stokes parameters~\cite{hkn97pre}.

The Lorentz group indeed gives an elegant formalism for polarization
optics.  Then what new physics does this generate?  This group has
an $E(2)$-like little group as we discussed in Sec.~\ref{max}, and
optical filters performing $E(2)$-like transformations may be
manufactured in laboratories~\cite{hkn97josa}.  In this case, those
$E(2)$-like filters will perform observable operations on light
waves which are mathematically equivalent to gauge transformations.

\section*{Acknowledgments}
Eugene Wigner was a son of Hungary.  The author is grateful to Prof.
Jozsef Janszky for inviting me to talk about this great physicist at
this meeting.  He would also like to thank Prof. Tamas Csorgo for
accompanying him when he was visiting Budapesti Evang\'elikas
Gimn\'azium where Eugene Wigner and John von Neumann studied in
their high-school years.


\begin{thebibliography}{99}

\bibitem{knp91}
Y. S. Kim and M. E. Noz, {\it Phase Space Picture of Quantum Mechanics}
(World Scientific, Singapore, 1991).

\bibitem{dir49}
P. A. M. Dirac, Rev. Mod. Phys. {\bf 21}, 392 (1949).

\bibitem{wig31}
E. P. Wigner, {\it Gruppentheorie und ihre Anwendung auf
Quantenmechanik der Atomspektren} (Vieweg, Braunschweig, 1931).

\bibitem{wig59}
E. P. Wigner, {\it Group Theory and its Application to the
Quantum Mechanics of Atomic Spectra}, translated from the German
by J. J. Griffin, Expanded and Improved Edition (Academic Press,
New York, 1959).

\bibitem{wig39}
E. P. Wigner,  Ann. Math. {\bf 40}, 149 (1939).

\bibitem{newig49}
T. D. Newton and E. P. Wigner, Rev. Mod. Phys. {\bf 21} 400 (1949).

\bibitem{hkn95}
D. Han, Y. S. Kim, and M. E. Noz, Phys. Lett. A {\bf 206}, 299 (1995).


\bibitem{stokes52}
G. G. Stokes, {\it On the composition and resolution of streams of
polarized light from different sources}, Trans. Cambridge Phil. Soc.
{\bf 9}, 399 (1852); Mathematical and Physical Papers (Cambrdige
Univ. Press) {\bf 3}, 233 (1902).

\bibitem{swind75}
W. Swindell, {\it Polarized Light} (Dowden, Hutchinson, and Ross, Inc.,
Stroudsburg, PA, 1975).

\bibitem{chou58}
K. C. Chou and L. G. Zastavenco, Zhur. Exptl. i Teoret. Fiz. 35,
1417 (1958) or Soviet Phys. JETP 8, 990 (1959).

\bibitem{jacob59}
M. Jacob and G. C. Wick, Ann. Phys. (New York) 40, 149 (1959).


\bibitem{hks85}
D. Han, Y. S. Kim, and D. Son, Unitary Transformations of Photon
Polarization Vectors, with D. Han and D. Son, Phys. Rev. D {\bf 31},
328 -- 330 (1985).

\bibitem{wein64}
S. Weinberg,  Phys. Rev. {\bf 134}, B882 (1964); {\it ibid.} {\bf 135},
B1049 (1964).

\bibitem{janner71}
A. Janner and T. Janssen,  Physica {\bf 53}, 1 (1971);
{\it ibid.} {\bf 60}, 292 (1972).

\bibitem{kuper76}
J. Kuperzstych, Nuovo Cimento {\bf 31B}, 1 (1976);
D. Han and Y. S. Kim,  Am. J. Phys. {\bf 49}, 348 (1981);
J. J. van der Bij, H. van Dam, and Y. J. Ng, Physica {\bf 116A},
307 (1982).

\bibitem{hks82}
D. Han, Y. S. Kim, and D. Son, Phys. Rev. D {\bf 26}, 3717 (1982).

\bibitem{hks86}
D. Han, Y. S. Kim, and D. Son,  Am. J. Phys. {\bf 54}, 818 (1986).

\bibitem{knp86}
Y. S. Kim and M. E. Noz, {\it Theory and Applications of the Poincar\'e
Group} (Reidel, Dordrecht, 1986).

\bibitem{baskal97}
S. Baskal and Y. S. Kim, Europhys. Lett. {\bf 40}, 375 (1997).

\bibitem{hkn96}
D. Han, Y. S. Kim, and M. E. Noz, Phys. Lett. A {\bf 206}, 299 (1996).

\bibitem{hkn97pre}
D. Han, Y. S. Kim, and M. E. Noz, Phys. Rev. E {\bf 56}, 6065 (1997).

\bibitem{hkn97josa}
D. Han, Y. S. Kim, and M. E. Noz, J. Opt. Soc. Am. A {\bf 14},
2290 (1997).

\end{thebibliography}
\end{document}